\documentclass{ws-ijmpcs}

\usepackage{hyperref}

\def\trimmarks{}

\begin{document}

\markboth{Granados and Weiss}
{Transverse charge and magnetization densities\ldots}

%
%

\title{TRANSVERSE CHARGE AND MAGNETIZATION DENSITIES IN THE NUCLEON'S 
CHIRAL PERIPHERY\footnote{Proceedings of 
QCD Evolution Workshop, Jefferson Lab, May 6--10, 2013,
\url{http://www.jlab.org/conferences/qcd2013/}}}

\author{C.~GRANADOS\footnote{Now at
Department of Physics and Astronomy, Nuclear Physics,
Uppsala University, Box 516, 75120 Uppsala, Sweden}, \; C.~WEISS}
\address{Theory Center, Jefferson Lab, Newport News, VA 23606, USA}

\maketitle


\begin{abstract}
In the light--front description of nucleon structure the
electromagnetic form factors are expressed in terms of frame--independent 
transverse densities of charge and magnetization. Recent work has studied 
the transverse densities at peripheral distances $b = O(M_\pi^{-1})$,
where they are governed by universal chiral dynamics and can be
computed in a model--independent manner. Of particular interest is
the comparison of the peripheral charge and magnetization densities.
We summarize (a) their interpretation as spin--independent and 
--dependent current matrix elements; (b) the leading--order chiral 
effective field theory results; (c) their mechanical interpretation
in the light--front formulation; (d) the large--$N_c$ 
limit of QCD and the role of $\Delta$ intermediate states; 
(e) the connection with generalized parton distributions and 
peripheral high--energy scattering processes.
\keywords{Generalized parton distributions,
transverse charge and magnetization densities, 
chiral effective field theory, light--front quantization}
\end{abstract}

\ccode{PACS numbers: 
11.15.Pg, 11.55.Fv, 12.39.Fe, 13.40.Gp, 13.60.Hb, 14.20.Dh
\\ Report number: JLAB-THY-13-1807}
\section{Transverse charge and magnetization densities}	
In the light--front description of relativistic systems the transition 
matrix elements of current operators are expressed in terms of transverse 
densities.\cite{Soper:1976jc,Miller:2007uy} 
They are defined as two--dimensional Fourier transforms of the 
invariant form factors and describe the distribution of charge and 
current in the system in transverse space. They are frame--independent 
(boost-invariant) and represent true densities in the light--front 
wave functions of composite systems, and thus provide an objective 
notion of the spatial structure of relativistic systems. 
Transverse densities have become an essential tool in the study of 
hadron structure, both in QCD and in approaches based on effective
degrees of freedom; see Ref.~[\refcite{Miller:2010nz}] for a review.

In the context of QCD, the transverse densities correspond to a projection 
(integral over the light--cone momentum fraction $x$) of the impact 
parameter--dependent parton densities,\cite{Burkardt:2000za} 
which are the transverse spatial
representation of the generalized parton distributions (GPDs). 
The transverse densities in the nucleon
thus provide indirect information on the 
distribution of partons in transverse space, and open an interesting 
connection between low--energy elastic $eN/\nu N$ scattering and 
high--energy inelastic processes resolving the nucleon's quark/gluon content.
Considerable efforts have been devoted to constructing empirical 
charge and magnetization densities in the nucleon from the available 
elastic form factor data.\cite{Miller:2010nz}

%
%
\begin{figure}
\parbox[c]{.26\textwidth}{
\includegraphics[width=.3\textwidth]{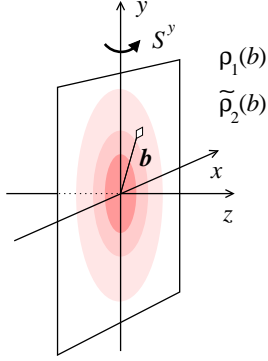}}
\vspace{-1ex}
\hspace{.15\textwidth}
\parbox[c]{.5\textwidth}{
\caption[]{Interpretation of the transverse densities associated
with the electromagnetic current, Eq.~(\ref{j_plus_rho}). 
$\rho_1(b)$ describes the spin--independent 
part of the $J^+$ current in a nucleon state with its transverse
center--of--momentum localized at the origin;
$\widetilde\rho_2(b)$, Eq.~(\ref{rho_2_tilde_def}), describes
the spin--dependent part in a nucleon polarized in the positive 
$y$--direction.}
\label{fig:interpretation}}
\end{figure}
The nucleon electromagnetic current matrix element is described by
the transverse charge and magnetization densities, defined as the Fourier 
transforms of the Dirac and Pauli form factors, $F_1(t)$ and
$F_2(t)$, in a frame where the momentum 
transfer to the nucleon is in the transverse direction,
\begin{equation}
\rho_{1, 2} (b) \;\; = \;\; \int \frac{d^2 \Delta_T}{(2\pi)^2} \; 
e^{-i \bm{\Delta}_T \bm{b}} \; F_{1, 2}(t = -\bm{\Delta}_T^2) .
\label{rho_def}
\end{equation}
Their integral over transverse space gives the total charge and
anomalous magnetic moment. The interpretation of $\rho_{1, 2}(b)$ 
as spatial densities has been discussed extensively in the 
literature.\cite{Miller:2007uy,Miller:2010nz,Burkardt:2000za} 
In a state where the nucleon is localized in transverse space at 
the origin, and polarized in the $y$--direction, the matrix element 
of the ``plus'' component of the current, $J^+ \equiv J^0 + J^z$,
at $x^\pm = 0$ and $\bm{x}_T = \bm{b}$, is given by
\begin{eqnarray}
\langle J^+ (\bm{b}) \rangle
&=& (...) \; \left[
\rho_1 (b) \;\; + \;\; (2 S^y) \, \cos\phi \, \widetilde\rho_2 (b) \right] ,
\label{j_plus_rho}
\\[2ex]
\widetilde\rho_2 (b) &\equiv& \frac{\partial}{\partial b} 
\left[ \frac{\rho_2(b)}{2 M_N} \right] ,
\label{rho_2_tilde_def}
\end{eqnarray}
where $(...)$ hides a trivial factor reflecting the normalization of states,
$\cos\phi \equiv b^x/b$ is the cosine of the azimuthal angle, 
and $S^y = \pm 1/2$ the spin projection in the $y$--direction in the nucleon
rest frame. The function $\rho_1(b)$ describes the spin--independent 
part of the current, the function $\widetilde\rho_2 (b)$ the 
spin--dependent part in a transversely polarized nucleon
(see Fig.~\ref{fig:interpretation}).
The latter changes sign between positive and negative values of $b^x$
(``right'' and ``left,'' when looking at the nucleon in the $z$--direction
from $+\infty$), as would be expected for a convection current due to 
rotational motion around the $y$--axis. A basic question of nucleon 
structure is how the ratio $\widetilde\rho_2 (b)/\rho_1 (b)$ behaves
as a function of the transverse position $b$, particularly in regions 
where a simple dynamical interpretation is possible. Intuition
from non--relativistic systems suggests that $\rho_1$ counts the number 
of constituents per transverse area, while $\widetilde\rho_2 (b)$ measures 
the current, so that the ratio should reflect the velocity of the
internal motion of the constituents. The concept of transverse densities 
allows one to pose the question rigorously also for relativistic systems.

At large distances the behavior of strong interactions is dominated
by the spontaneous breaking of chiral symmetry. The associated 
quasi--massless excitations (Goldstone bosons), the pions,
couple weakly to hadronic matter and mediate low--energy interactions 
over distances of the order $M_\pi^{-1}$, much larger than the bulk
hadronic size. In current matrix elements they induce characteristic
contributions in which the current couples to the hadron by exchange
of ``soft'' pions with momenta $k_\pi \sim M_\pi$. In the transverse
densities they give rise to distinctive long--range components at
$b = O(M_\pi^{-1})$, which can be calculated model--independently
and represent fundamental chiral properties of the structure. They are 
analogous to the well--known ``Yukawa tails'' of non--relativistic physics
but have a precise relativistic meaning. Recent theoretical 
work\cite{Granados:2013moa,Strikman:2010pu} 
has studied the chiral component of the transverse charge and 
magnetization densities using methods of chiral effective 
field theory (EFT). It observed an interesting 
inequality\cite{Granados:2013moa} 
between the leading--order spin--dependent and independent peripheral 
densities, and suggested a simple explanation in a mechanical picture 
based on the first--quantized light--front formulation of chiral EFT.
It also investigated the scaling behavior of the transverse densities
in the large--$N_c$ limit of QCD and showed that inclusion of
$\Delta$ intermediate states guarantees the proper scaling of the pionic 
component. These findings provide model--independent constraints on the 
peripheral transverse densities at large $b$ and enable an intuitive 
understanding of chiral nucleon structure. Here we summarize the
main points, focusing on the comparison of $\widetilde\rho_2$ and
$\rho_1$; for the conceptual and practical aspects of chiral EFT
in peripheral transverse densities we refer to the 
original article.\cite{Granados:2013moa} 
\section{Peripheral densities from chiral EFT}
\begin{figure}
\begin{center}
\includegraphics[width=.9\textwidth]{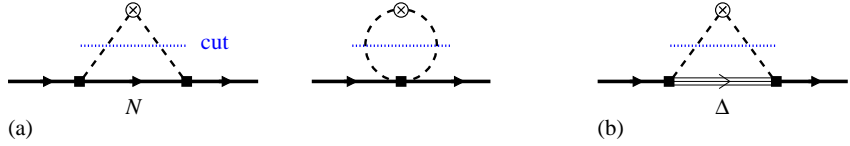}
\vspace{-1ex}
\end{center}
\caption[]{(a) LO chiral EFT processes contributing to the 
two--pion cut of the isovector nucleon form factors and the
peripheral charge and magnetization densities. (b) Intermediate
$\Delta$ contribution.}
\label{fig:diag}
\end{figure}

\begin{figure}
\parbox[c]{.6\textwidth}
{\includegraphics[width=.6\textwidth]{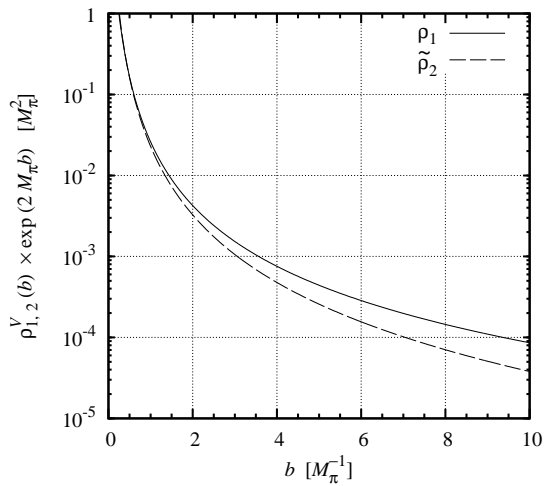}}
\vspace{-1ex}
\hspace{.03\textwidth}
\parbox[c]{.32\textwidth}
{\caption[]{LO chiral component of the nucleon's isovector 
transverse spin--independent density $\rho_1^V(b)$ 
(solid line) and spin--dependent density 
$\widetilde \rho_2^V(b)$ (dashed line).\cite{Granados:2013moa}  
The plot shows the densities
with the exponential factor $\exp (- 2 M_\pi b)$ extracted
[the functions plotted correspond to the pre-exponential factor in 
Eq.~(\ref{large_b_general})]. The distance $b$ is given in units
of $M_\pi^{-1}$, the densities in units of $M_\pi^2$.}
\label{fig:rho_current}}
\end{figure}

The large--distance behavior of transverse densities can conveniently
be studied in a dispersive representation,\cite{Strikman:2010pu} 
in which they are expressed
as integrals over the imaginary parts (or spectral functions) of the
invariant form factors on the principal cut in the timelike region,
\begin{equation}
\rho_{1, 2} (b) \;\; = \;\; \int\limits_{4M_\pi^2}^\infty \!\frac{dt}{2\pi} 
\; K_0(\sqrt{t} b) \; \frac{\textrm{Im}\, F_{1, 2} (t + i0)}{\pi} .
\label{rho_dispersion}
\end{equation}
In this formulation the densities as $b = O(M_\pi^{-1})$ are the ``image'' 
of the spectral functions on the two--pion cut at $t - 4 M_\pi^2 =
O(M_\pi^2)$, which can systematically be computed in chiral EFT and
have been studied extensively in the literature.
\cite{Gasser:1987rb,Bernard:1996cc,Kubis:2000zd,Kaiser:2003qp}
In the leading--order
(LO) approximation of relativistic chiral EFT,\cite{Becher:1999he} 
the isovector spectral 
functions [$V \equiv (p - n)/2$] are given by the chiral processes of
Fig.~\ref{fig:diag}a, where the current couples to the nucleon by
two--pion exchange, and the pion--nucleon vertices are those of the LO
chiral Lagrangian. Evaluation of the chiral component of the densities
is straightforward, and analytic approximations can be derived.
The general form of the peripheral densities at $b = O(M_\pi^{-1})$ is
%
%
\begin{equation}
\rho_1^V (b), \; \widetilde\rho_2^V (b) 
\; = \; e^{-2 M_\pi b} \times \textrm{functions} \, (M_\pi, M_N; b) ,
\label{large_b_general}
\end{equation}
where the exponential behavior is dictated by the minimum mass of the 
exchanged two--pion system in the $t$--channel and and the pre--exponential 
functions reflect the complexity of its coupling to the nucleon.
In leading order of $M_\pi/M_N$ (heavy--baryon expansion)
one finds that
\begin{equation}
\widetilde\rho_2^V (b) / \rho_1^V (b) \;\; = \;\; 
O[ (M_\pi/M_N)^0 ] \;\;  \equiv \;\; O(1)
\hspace{2em} [b = O(M_\pi^{-1})] .
\label{rho_2_tilde_order}
\end{equation}
The spin--independent and --dependent parts of the isovector current 
matrix element Eq.~(\ref{rho_2_tilde_def}) are of the same order in the
chiral region. This shows that it is natural to work with 
$\widetilde\rho_2$ instead of $\rho_2$; for the original
$\rho_2$ the corresponding ratio is $\rho_2/\rho_1 = O(M_N/M_\pi)$.
The numerical results for the LO densities in the chiral region
(see Fig.~\ref{fig:rho_current}) show that in this approximation 
the spin--dependent density is bounded by the 
spin--independent one,
\begin{equation}
|\widetilde\rho_2^V (b)| \; < \; \rho_1^V (b) .
\label{inequality}
\end{equation}
The inequality is practically saturated at distances 
$b \sim \textrm{few}\, M_\pi^{-1}$; at larger distances the 
spin--dependent density becomes systematically smaller than 
the spin--independent one. We emphasize that Eq.~(\ref{inequality}) 
is specific to the LO chiral EFT result and can be modified by higher--order
corrections;\cite{Kubis:2000zd,Kaiser:2003qp} it also ceases to
be valid when $\Delta$ intermediate states are included as required
by the large--$N_c$ limit of QCD (see below). Still, it is interesting 
to inquire about the dynamical origin of this surprising observation.
%
%
\section{Mechanical picture in light--front quantization}
\begin{figure}
\begin{center}
\includegraphics[width=.95\textwidth]{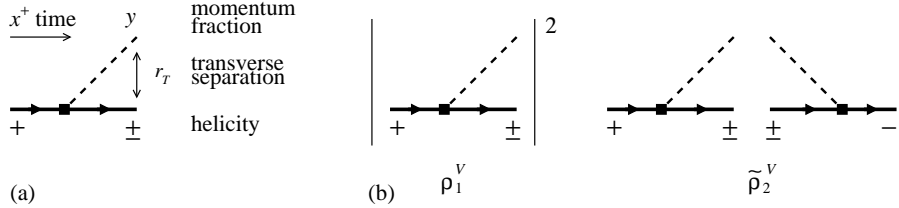}
\vspace{-2ex}
\end{center}
\caption[]{(a) Light--front wave function of the $\pi N$ system
in chiral EFT. (b) Wave function overlap representation of the
transverse densities $\rho_1$ (helicity--conserving)
and $\widetilde\rho_2$ (helicity--flip).}
\label{fig:diag_wf}
\end{figure}

A more intuitive understanding of the structure of the LO chiral component,
can be obtained in a first--quantized mechanical picture, where one follows 
the evolution of chiral EFT processes in light--front time $x^+$.\cite{inprep} 
In this formulation the 
peripheral densities arise from processes in which the initial nucleon 
``fluctuates'' into a large--size pion--nucleon ($\pi N$) system through 
the perturbative chiral EFT interaction. The process is described by 
light--front wave functions $\psi_{0, 1}(y, r_T)$
(see Fig.~\ref{fig:diag_wf}a), which depend on
the pion light--cone momentum fraction $y = O(M_\pi / M_N)$,
the transverse spatial separation $r_T = O(M_\pi^{-1})$ of the 
final $\pi N$ system, and the light--front helicities of the initial 
and final nucleon (helicity--conserving, $\psi_0$; helicity--flip,
$\psi_1$; the dependence on the transverse angle $\phi$ is
dictated by angular momentum conservation and can 
be separated\cite{inprep}). The peripheral isovector densities can 
then be expressed as overlap integrals 
of the light--front wave functions (see Fig.~\ref{fig:diag_wf}b),
\begin{equation}
\left.
\begin{array}{l}
\rho_1^V (b)
\\[2ex]
\widetilde\rho_2^V (b)
\end{array} 
\right\}
\;\; = \;\; \int\!\frac{dy}{2\pi y \bar y^3}\;
\left\{
\begin{array}{l}
[ |\psi_{0}(y, r_T)|^2 + |\psi_{1}(y, r_T)|^2 ]_{r_T = b/\bar y} 
\; + \; \textrm{instant.,}
\\[2ex]
[ \psi_{0}^\dagger (y, r_T) \; \psi_{1}(y, r_T)
+ \psi_{1}^\dagger (y, r_T) \; \psi_{0}(y, r_T) ]_{r_T = b/\bar y} ,
\end{array}
\right.
\label{overlap}
\end{equation}
where $\bar y \equiv 1 - y$. The inequality Eq.~(\ref{inequality})
follows directly from the quadratic forms in the integrand of 
Eq.~(\ref{overlap}). It holds up to an instantaneous term 
$\propto \delta(y)$ in $\rho_1$, which represents the cumulative 
effect of high--mass intermediate states not resolved in chiral
EFT; this term is proportional to $(1 - g_A^2)$ and 
numerically small.\cite{Granados:2013moa} 
Likewise, Eq.~(\ref{rho_2_tilde_order}) can be
inferred from the parametric order of the wave functions.\cite{inprep} 
These results show that the light--front formulation of chiral EFT provides 
genuine new insight into chiral nucleon structure and demonstrates
the usefulness of the transverse densities as quantities with a
direct interpretation in terms of light--front wave functions.
%
%

Alternatively, one may characterize the nucleon polarization in
Eq.~(\ref{overlap}) by its transverse spin in the rest frame.
This leads to an even simpler mechanical 
picture:\cite{Granados:2013moa} the nucleon with transverse
spin projection $S^y = +1/2$ fluctuates into a $\pi N$ system, where the
intermediate $N$ has spin projection $-1/2$ and the $\pi$ orbits
with $L^y = +1$ around the $y$ axis. The ratio of spin--dependent 
and --independent peripheral densities is of the order of the pion 
velocity, $v = k_\pi/M_\pi = O(1)$, which naturally explains 
Eq.~(\ref{rho_2_tilde_order}). This picture is maximally
close to the non--relativistic intuition and takes advantage
of the fact that the light--front description is boost--invariant
and can be implemented in any frame, including the rest frame.
Details will be provided elsewhere.\cite{inprep} 
\section{$\Delta$ intermediate states and large--$N_c$ limit}
Other interesting information on the peripheral transverse densities
comes from the large--$N_c$ limit of QCD.\cite{Granados:2013moa} 
The $N_c$--scaling of
integral nucleon observables (charges, magnetic moments) has been
studied extensively in the literature\cite{Witten:1979kh,Dashen:1993jt} 
and provides useful constraints
for dynamical models and phenomenological analysis. The same
techniques can be applied to study the transverse densities at
non--exceptional distances $b = O(N_c^0)$, which includes the
chiral region $b = O(M_\pi^{-1})$, as $M_\pi = O(N_c^0)$. 
One finds that the isovector densities in large--$N_c$ QCD scale as
\begin{equation}
\rho_1^V (b) \;\; = \;\; O(N_c^0),
\hspace{2em}
\widetilde \rho_2^V(b) \;\; = \;\; O(N_c)
\hspace{2em}
[b = O(N_c^0)].
\label{rho_largen_general}
\end{equation}
The spin--dependent density is parametrically larger than the 
spin--independent one, reflecting a general property of the
spin--flavor wave function of the large--$N_c$ nucleon. 
It implies that the inequality Eq.~(\ref{inequality}), 
observed in the LO chiral EFT result, is not compatible with the 
large--$N_c$ limit of QCD. Indeed, one finds that the spin--independent 
density obtained in LO chiral EFT exhibits a ``wrong'' $N_c$--scaling 
as $\rho_1^V (b) = O(N_c)$.

This is not surprising, as it is well--known that in the large--$N_c$
limit the $\Delta$ isobar becomes degenerate with the $N$ and needs
to be included as an intermediate state in the two--pion spectral
function on the same footing 
(see Fig.~\ref{fig:diag}b).\cite{Cohen:1992uy,Cohen:1996zz}
Using a phenomenological $\pi N \Delta$
coupling it was shown explicitly\cite{Granados:2013moa} 
that in the large--$N_c$ limit the intermediate $\Delta$
contribution \textit{cancels} the $N$ one in $\rho_1(b)$ to leading order
in $1/N_c$, while it \textit{adds} to the $N$ in $\widetilde \rho_2^V(b)$.
Inclusion of the $\Delta$ thus ``restores'' the proper $N_c$--scaling
of the two--pion component of the peripheral transverse charge density,
and one obtains 
\begin{eqnarray}
\rho_1^V (b)_N \; + \; \rho_1^V (b)_\Delta &=& O(N_c^{0}) ,
\label{largen_rho1}
\\[0ex]
\widetilde\rho_2^V (b)_N 
\; + \; \widetilde\rho_2^V (b)_\Delta &=& O(N_c) 
\hspace{2em} [b = O(N_c^0)] ,
\label{largen_rho2_tilde}
\end{eqnarray}
in accordance with Eq.~(\ref{rho_largen_general}).
The $\Delta$ thus plays an essential qualitative role in the large--$N_c$
limit. Numerically, with the empirical values of the $N$-$\Delta$ mass 
splitting and $\pi N \Delta$ coupling, the $\Delta$ contribution to the 
peripheral densities is $\lesssim 20\%$ of the $N$ at $b > 2\, M_\pi^{-1}$
in both cases, so that the chiral EFT result with $N$ only still
provides a valid approximation.

The fact that the spin--dependent component of the isovector transverse 
density is parametrically leading in the large--$N_c$ limit suggests that 
it may also be numerically larger than the spin--independent one.
This question could be addressed in dynamical models based on the
large--$N_c$ limit, which describe the nucleon as a chiral soliton
(skyrmion,\cite{Zahed:1986qz} chiral quark--soliton 
model\cite{Christov:1995vm}). In these models the $N$
and $\Delta$ correspond to different rotational states of the 
classical soliton, and the contribution from $\Delta$ intermediate 
states is effectively included in the peripheral 
densities.\cite{Cohen:1992uy,Cohen:1996zz}
\section{Empirical densities and experimental tests}
Chiral dynamics and the large--$N_c$ limit of QCD provide model--independent
theoretical insight into the structure of the nucleon's peripheral
transverse charge and magnetization (or spin--independent and dependent)
densities. Important practical questions are how one should construct
empirical densities incorporating these constraints, and whether one
could probe the chiral component of the transverse densities directly
in scattering experiments.

A numerical study of the nucleon's transverse 
densities\cite{Miller:2011du} using the dispersion 
integral Eq.~(\ref{rho_dispersion}) 
shows that the chiral two--pion component becomes numerically dominant 
only at large distances $b \gtrsim 2\, \textrm{fm}$. At smaller distances the 
densities (isovector and isoscalar) are overwhelmingly generated by the 
vector meson resonances ($\rho, \omega$) in the spectral function.
Realistic transverse densities are therefore best constructed by 
evaluating the 
dispersion integral Eq.~(\ref{rho_dispersion}) with empirical
spectral functions, which include the chiral two--pion component 
near threshold (as described by chiral EFT), 
the vector meson resonances, and
a continuum of higher--mass states determined by fits to the spacelike
form factor data.\cite{Belushkin:2006qa,Lorenz:2012tm} This formulation 
implements the proper analytic structure of the form 
factor near threshold, which guarantees the correct large--distance 
asymptotic behavior of the densities.

The chiral component at large $b$ makes a distinctive contribution 
to the $b^{2n}$--weighted moments ($n = 1, 2, \ldots$) of the 
transverse densities, which are related to the $n$'th derivatives 
of the elastic form factors at 
$t = 0$.\cite{Granados:2013moa} A preliminary assessment\cite{Strikman:2010pu} 
found that the chiral component of the isovector
density contributes only $\sim 20\%$ to the $b^2$--weighted moment,
but $O(1)$ to the $b^4$--weighted and higher moments. This suggests 
that the chiral component causes ``unnatural'' behavior of the second 
and higher derivatives of the
form factors at $t = 0$, which might be observable in combined fits
to low--$|t|$ elastic scattering data and atomic physics measurements
of the proton charge radius.\cite{Bernauer:2013tpr,E12-11-106}

The chiral component of transverse nucleon structure can also be probed
in peripheral high--energy scattering processes, e.g.\ deep--inelastic
or exclusive processes which resolve the peripheral quark/gluon structure
with a momentum transfer $Q^2 \gg 1\, \textrm{GeV}^2$. An important point
is that the chiral light--front wave functions of the peripheral $\pi N$ 
system are universal and determine also the nucleon's peripheral 
quark/gluon densities,\cite{Strikman:2009bd} or the amplitudes for 
peripheral exclusive 
processes with pion production, when supplemented with the appropriate
information about pion structure. One promising candidate is hard
exclusive electroproduction (of vector mesons, photons etc.) on a 
peripheral pion that is observed in the final state together with the
remnant nucleon;\cite{Strikman:2003gz} 
such types of processes could be studied with
a future Electron--Ion Collider 
(EIC).\cite{Accardi:2011mz,Accardi:2012hwp} The advantage of the 
light--front formulation of nucleon structure is precisely that it 
connects low--energy elastic form factors with observables in 
peripheral high--energy scattering processes in a well--defined 
manner, greatly enlarging the number of available probes.
\\[2ex]
{\small {\bf Notice:} Authored by Jefferson Science Associates, 
LLC under U.S.\ DOE Contract No.~DE-AC05-06OR23177. The U.S.\ Government 
retains a non--exclusive, paid--up, irrevocable, world--wide license to 
publish or reproduce this manuscript for U.S.\ Government purposes.}
\end{document}